\documentstyle[emulateapj]{article}
\input{epsf} 
\def \etal      {et al.\ }

\def \fgas {f_{\rm gas}}

\def \msun {\hbox{${\rm M}_{\sun}$}}

\def
\betamodel {\hbox{$\beta$--model} }
\def \betamodels  {\hbox{$\beta$--models} }

\def \fgas {\hbox{$f_{gas}$}}

\def \kev       {{\rm\ keV}}

\begin{document}

\title{Is MS1054-03 an exceptional cluster? A new investigation of ROSAT/HRI
X-ray data}

\author{ D.M. Neumann , M. Arnaud}
\affil{
CEA/Saclay,
Service d'Astrophysique, Orme des
Merisiers, B\^at. 709,
91191
Gif-sur-Yvette Cedex, France
email:ddon@cea.fr ; arnaud@hep.saclay.cea.fr
}

\begin{abstract}
We reanalyzed the ROSAT/HRI observation of MS1054-03, optimizing the
channel HRI selection and including a new exposure of 68 ksec.
From a wavelet analysis of the HRI image we identify the main cluster
component and find evidence for substructure in the west, which might
either be a group of galaxies falling onto the cluster or a foreground
source.

Our 1--D and 2--D analysis of the  data show that the cluster can
 be fitted well by a classical \betamodel centered only $20\arcsec$
away from the central cD galaxy.  The core radius and $\beta$ values
derived from the spherical model($\beta = 0.96_{-0.22}^{+0.48}$) and
the elliptical model ($\beta = 0.73\pm0.18$) are consistent.

We derived the gas mass and total mass of the cluster from the
\betamodel fit and the previously published ASCA temperature
($12.3^{+3.1}_{-2.2} \kev$).  The gas mass fraction at the virial
radius is $\fgas = (14\small{[-3,+2.5]} \pm 3\ ) \%$ for $\Omega_0=1$,
where the errors in brackets come from the uncertainty on the
temperature and the remaining errors from the HRI imaging data.  
The gas mass 
fraction computed for the best fit ASCA
temperature is significantly lower than found for nearby hot clusters,
$\fgas = 20.1\ \pm 1.6\ \%$.  This local value can be matched if the
actual virial temperature of MS1054-032 were close to the lower ASCA limit 
($\sim
10~\kev$) with an even lower value of $8~\kev$ giving the best agreement.
Such a bias between the virial and measured temperature could be due to
the presence of shock waves in the intracluster medium stemming from recent 
mergers.  Another
possibility, that reconciles a high temperature with the local gas
mass fraction, is the existence of a non zero cosmological constant.

\end{abstract}


\section{Introduction}

MS1054-03 is the most distant cluster of galaxies found in the X-ray
selected sample of the {\it Einstein} Extended Medium Sensitivity
Survey (EMSS; Gioia et al., 1990).  Due to the extreme sensitivity of
the high--mass end of the cluster mass function to the density
parameter, $\Omega_0$, the existence of massive, virialized clusters at
high z is of great cosmological significance (e.g.  Oukbir \& Blanchard
1992).  The very existence of even a few massive clusters at redshifts
approaching unity and beyond strongly argue against cosmological models with
$\Omega_0=1$.

There is a general consensus that MS1054-03 is such a massive cluster,
based on its high X-ray temperature, $kT=12.3^{+3.1}_{-2.2}$~keV,
measured with ASCA (Donahue et al.  1998 hereafter D98), its high
velocity dispersion (Tran \etal 1999) and intense weak lensing signal
(Luppino \& Kaiser 1997).  On the other hand its apparent complex morphology
(D98) might indicate that MS1054-03 has not yet reached an equilibrium
state, which could cast some doubt on the dynamical mass estimates and
on the relevance of the cluster for cosmological tests.  Tran \etal
(1999) emphasized the good agreement between the various mass
estimates, but the large error bars must be noted.

In this paper we re-investigate the ROSAT/HRI observations of MS1054,
including a new exposure taken in 1997.  We first try to better
understand the cluster morphology.  A wavelet analysis of the HRI
image is performed to unravel significant substructures and identify
the main cluster component.  A $\beta$--model is then fitted to the
data.  From the fit results, we estimate for the first time the gas
mass of this cluster.  A comparison of the gas mass fraction of
MS1054-032, with the gas mass fraction of nearby clusters, provides a
consistency check on the total mass estimate, assuming that the gas
mass fraction in clusters does not evolve with redshift.

Throughout the paper we assume $H_0 = 50$ km/sec/Mpc, $q_0=0.5$,
$\Lambda=0$ and all quoted error bars are 1 $\sigma$ unless otherwise
stated.  At the redshift of the cluster, $z=0.83$, one arcmin corresponds to 
$497{\rm h}_{50}^{-1}$~kpc.

\section{Observations}

MS1054-03 was observed by the ROSAT HRI (Tr\"umper 1992) for
190,754~sec in total: an exposure of $\sim122$~ksec in 1996, which was
analyzed and presented in D98 and a pointing of 68~ksec taken in 1997.
In this analysis we combined the exposures of 1996 and 1997 and we
selected only HRI channels 1 to 7 (David et al.  1997) in order to
maximize the signal-to-noise ratio.  The background level is thus
lowered by a factor of about 2.  The HRI image smoothed with a Gauss
filter is shown on Fig.~1 and looks similar to the image presented by
D98.



\section{Morphology}

In order to remove noise and to identify the significant structures we
performed a wavelet-analysis of the HRI raw image with the Multi-scale
Vision Model (MVM) package (Ru\'e \& Bijaoui 1997).  We explicitly
took into account the Poisson statistics and the significance level
was set to $3\sigma$.  The algorithm is optimized for the detection of
diffuse components by normalizing the wavelet coefficients by their
energy (Ru\'e \& Bijaoui 1997, Arnaud \etal 2000).

The reconstructed image consists of point sources 
and one large scale diffuse component at the cluster position.
To extract possible substructures, we reapplied  the multi-scale
analysis to the reconstructed diffuse source only.  Two components are
detected that are shown in Fig.~2, superimposed on the cluster optical
image.  The main component ($\sim 90\%$ of the flux) is centered 
at RA = $10^h56^m58.47^s$, Dec.= $-03\deg37\arcmin31.2\arcsec$, 
about $20~\arcsec$ ($166~{\rm h}_{50}^{-1}$~kpc)
from the dominant galaxy (D98).  We identify this structure
with the main cluster component.  It is very elliptical with an
elongation in the east-west direction.  The orientation follows the
filamentary structure found in the optical and in the weak lensing map
(Luppino \& Kaiser 1997).  The second component is compact and lies in
the west, with an offset of 0.6 arcmin with respect to the main
cluster center.  It coincides with the brightest image peak,
identified by D98, but its reconstructed flux is only $\sim 10\%$ of
the main cluster flux.

This western source might be a subcluster gravitationally connected to
the main cluster, or a projection effect due to a group of galaxies in
the field-of-view (FOV) or even a point source.  The available data are 
insufficient
to settle this issue.  There is a weak indication that the source is
extended, from our comparison of the source surface brightness profile
with the point spread function (PSF) of the ROSAT/HRI (David et al. 1997). 
However, due to the 
lack of bright
sources in the FOV, we cannot correct for possible alignment
uncertainties, and the extent might be due to errors in
the aspect reconstruction
 or residual
main cluster emission.  Very close to the maximum of this source we
find several cluster members so this structure might  be a
subgroup falling right onto the cluster.  Finally it must be noticed
that the reconstructed surface mass density by Luppino \& Kaiser
(1997) presents an extension in the west.  Although it is unresolved,
it might indicate that the western X--ray component does trace a
gravitational potential.

\section{Isothermal beta-model fits}

%

We performed $\beta$-model fits to the data using
spherically symmetric (1--D) and elliptical (2--D)
\betamodel fits.  We excluded from the fits circular
regions corresponding to serendipitous point sources. In the 1--D case
the X-ray surface brightness profile is given by:
\begin{equation}
S = S_0 \left(1 + r^2/r_c^2\right)^{-3\beta+0.5} +B
\end{equation}
where $r_c$ is the core radius, $\beta$ is a slope parameter and
$B$ is the background level.
The functional form for the elliptical \betamodel  can be found in Neumann \&
B\"ohringer (1997).

\subsection{Spherical model}

We binned the data in concentric annuli with the center at
the position determined from the above wavelet reconstruction.  The
\betamodel was convolved with the PSF of the
ROSAT/HRI. The free parameters in the fit are $S_0$, $r_c$, $\beta$ and $B$.
We first consider the main cluster component in a common way by excluding 
non-symmetric features in the data, i.e. a circular region around the 
substructure in the west.
 Fig.~3 shows the observed surface
brightness profile and the best fit model.  The reduced $\chi ^2$
value is $\chi ^2 = 0.97$, and the shape parameters are well
constrained (see Tab.\ref{tab:beta2}).  On the other hand if the
substructure is not removed, a good fit is still obtained ($\chi ^2 =
0.99$) but the best fit shape parameters are unreasonably large
($\beta=2., r_{c}= 840~{\rm kpc}$), and poorly constrained
($\beta>1.1$).  An artificial increase of the best
fit shape parameters is a well known effect of not excising a secondary
sub-cluster in the $\beta$--model fit (Jones \& Forman 1999).  A good fit is 
still
obtained due to the large errors on the observed profile, whereas no
upper limit on $\beta$ can be set because the best fit core radius
becomes similar to the maximum radius of cluster emission detected.

\subsection{Elliptical model}

We also fitted an elliptical isothermal \betamodel to the HRI image.
Again we excise the substructure in
the west as well as other weak unresovled sources in the FOV. 
  Due to the limited statistics of the observation, we did not
attempt to fit the whole image with a \betamodel and an additional
component for the substructure, since the elliptical \betamodel has already 8 
fit parameters.

The iso-contours of the best fit model are
superimposed on the cluster image in Fig.~1.  The fitting procedure
is described in full detail in Neumann (1999).  We binned the data
into an image with a pixel size of $5\arcsec \times 5\arcsec$.  The fit
included 
all pixels less than 4.6~arcmin away from the central pointing
position of $10^h57^m00.00^s$, and Dec.= $-03\deg37\arcmin12.0\arcsec$
(J2000).  The central position of the cluster was left free to vary.

As we are dealing with low number statistics in each image pixel,
which shows non-Gaussian behavior, we smoothed the data with a Gauss
filter ($\sigma=10\arcsec$) before fitting.  The modeling accounts for
the Gauss filtering and for the PSF (the fitted model is convolved with the
Gauss filtered PSF). To calculate the errors of the
\betamodel parameters we performed a Monte-Carlo analysis in which we
added Poisson noise to the data and fitted the \betamodel
subsequently.  We performed 100 Monte-Carlo realizations.

The validity of this approach was discussed in Neumann \& B\"ohringer
(1997) and Neumann (1999).  In order to see whether it is still free
of systematics in this regime of extremely low signal-to-noise data
(the central cluster intensity is only a factor of two higher than the
background level) we simulated 100 realizations of the image of the
best fit \betamodel cluster including background.  The Poisson
statistics were defined according to the length of the actual exposure
time.  We subsequently fitted a \betamodel to these artificial
smoothed images, as for the real image.  The results are satisfactory,
as the mean output values are practically identical to the input
values, with differences much smaller than the actual determined error
bars for MS1054-03.  The width of their distribution is slightly
smaller than the errors determined from the real data.

\subsection{Comparison of the 1--D and 2--D models}

The values of the spherical and elliptical \betamodel parameters for
the main cluster component together with their 1 $\sigma$ errors are
given in Tab.\ref{tab:beta2} and the two \betamodels profiles are
plotted on Fig.~3.  The best fit center of the elliptical model is
only $5~\arcsec$ away from the central position from the wavelet
analysis, which we also choose as the center for the 1--D fit.  As already
mentioned, these central positions are also in excellent agreement
with the position of the brightest cluster member.  The 1--D and 2--D 
best fit parameters are
consistent.  The background level is slightly higher in the 1--D fit
which explains the lower central value and the higher value for
$\beta$. Nevertheless the $68\%$ confidence intervals for the fitted
shape parameters show a
large large region of overlap.


\section{Luminosity}

The background estimated from the 1--D model fit was subtracted from
the radial profile.  An additional $5\%$ systematic uncertainty was
assumed on its level.  The cluster emission is detected up to
$2\arcmin$ or $1~{\rm h}_{50}^{-1}~{\rm Mpc}$ at the $68\%$ confidence
level.  The observed count rate within this aperture is $(9\pm
2)\times10^{-3}$ cts/s, $\sim 25\%$ higher than the value given by D98
(subtracting additional sources, including the western structure for
consistency).  We recall that the count rate of the western
substructure is only $10\%$ of the cluster flux.  The count rates
derived from the best fit 2--D
model is consistent but higher $11\times~10^{-3}$cts/s, a
consequence of the lower best fit background, in this case.

The observed count rate was translated into luminosity, assuming a
cluster temperature of 12.3~keV, as measured by D98 from ASCA data.
The derived X--ray luminosity is $L_{\rm X}=1.2\times10^{45} {\rm
h}_{50}^{-2}$ ergs/s (0.1--2.4 keV rest frame).  The bolometric
luminosity within $2\arcmin$ in radius, $L_{\rm bol}=4.3\pm0.9 \times
10^{45} {\rm h}_{50}^{-2}$ergs/s, is consistent with the ASCA value of
$4.4 \times 10^{45} {\rm h}_{50}^{-2}$ergs/s (D98).  This might be
fortuitous though since the later includes all components in the
field.  The bolometric luminosity is lower than the value
expected from the $L_{\rm X}$--$T$ relation of Arnaud \& Evrard
(1999), assuming this relation does not evolve with redshift.  For a
cluster with $T=12.3^{+3.1}_{-2.2}$~keV this correlation predicts a
luminosity of $9.1^{+8.2}_{-4.0} \times 10^{45}~{\rm
h}_{50}^{-2}$ergs/s.  Up to now it is not clear whether the $L_{\rm
X}$--$T$ relation evolves with redshift or not (see for example
Schindler 1999).  Also, the total X-ray luminosity of MS1054 might be
actually higher since the cluster might extend beyond the radius of
detection.  However, due to the low S/N ratio of this observation such
extrapolation is very sensitive to the assumed background level and
$\beta$ parameter.

\section{Mass content}

\subsection{Mass estimates}
\label{sec:mass}
The cluster gas mass profile $M_{\rm gas}(r)$ can be derived from the
best fit \betamodel, given the observed temperature and $N_{H}$
values.  As the emissivity in the ROSAT/HRI energy band is nearly
insensitive to the temperature, the uncertainty on the gas mass is 
overwhelmingly 
dominated by the errors on the \betamodel parameters. The
association of the western source with the cluster is unclear.
However, in practice, this additional uncertainty has no significant
impact on the
gas mass estimate.  If the substructure is included in the \betamodel
analysis, the derived gas mass differs by less than $5\%$ from the
value obtained when excising it; the derived mass is increased within
the detection radius (as a consequence of the $\sim 10\%$ higher flux)
and artificially decreased when one extrapolates the data to higher
radii (due to the higher derived $\beta$ value).  In the following we
only consider the main cluster component and its corresponding $\beta$
model.

The total mass can be estimated with the isothermal \betamodel approach
(BM), which is thought to be roughly valid even if the cluster is not
fully in hydrostatic equilibrium (e.g Schindler 1996).  The
alternative is to employ the virial theorem (VT) at given density
contrast, over the mean mass density of the Universe at the cluster redshift,
normalized from numerical simulations (Evrard \etal 1996).  The
resulting $M$--$T$ relation (see Appendix) depends both on redshift
and on the cosmological parameters $\Omega_0$ and $\Lambda$ (Voit \&
Donahue 1998).  We used the analytical expression from Bryan
\& Norman (1998).

We first fix the temperature to the best fit ASCA value of
k$T=12.3~\kev$ and consider an $\Omega_0=1$ Universe.  The uncertainty 
introduced
by the errors on the temperature is discussed later.  The statistical
errors on $M_{\rm gas}(r)$ and on $M_{\rm BM}(r)$ due to the
uncertainties in the \betamodel, are estimated following the general
method described in Elbaz \etal (1995).  The derived mass profiles are
plotted in Fig.~4.  Tab.2 summarizes the mass estimates at $1~{\rm
h}_{50}^{-1}~{\rm Mpc}$ (the maximum radius of detection) and
$1.65~{\rm h}_{50}^{-1}~{\rm Mpc}$ (the virial radius $R_{\rm V}$ for
z=0.83 and k$T=12.3~\kev$ using the Evrard \etal (1996) simulations).



The 2--D model gas mass at $1~{\rm h}_{50}^{-1}~{\rm Mpc}$ is $15\%$
higher than the value derived from the 1--D model.  This discrepancy,
larger than the formal statistical uncertainties, is essentially due to
the systematic difference in the background estimates, which 
dominates the error.  As the derived $\beta$ value is consistently
smaller in the 2--D fit, this discrepancy is amplified for extrapolated
gas masses.  It reaches $25\%$ at the virial radius.  Similarly the
total BM mass estimate which scales as $\beta$ is smaller for the 2--D
model than for the 1--D model.
Both are consistent with the VT estimate,
which we will adopt in the following discussion.  The gas mass is taken
as the average of the 1--D and 2--D estimates.  Their difference is an
estimate of the systematic uncertainties that we add in quadrature to
statistical ones.  The gas mass fraction at the virial radius is thus
$\fgas = 14 \pm 3 \%$, for k$T=12.3~\kev$ and $\Omega_0=1$, the error
coming from the uncertainty on the gas mass from the imaging data.

We performed the same analysis for the extreme values of the
temperature, as allowed by the ASCA data at the $90\%$ confidence
level (D98).
For the lower limit on the temperature, k$T=10.1~\kev$, the virial
radius is $1.58~{\rm Mpc}$ and the gas mass fraction within that
radius  is $\fgas = 16.5 \pm 3 \%$.  For k$T=15.4~\kev$, we get
$R_{\rm V}=1.95~{\rm Mpc}$ and $\fgas = 11 \pm 3 \%$.

In summary the gas mass fraction at the virial radius is $\fgas =
14\small{[-3,+2.5]} \pm 3\ \%$ for $\Omega_0=1$. We have separated the
uncertainties due to the errors on the temperature  (in bracket), which
affect essentially the total mass estimate,  and the
uncertainties from the imaging data, which only affect the gas mass
estimate.

\subsection{Comparison with nearby clusters}
\label{sec:compare}

As mentioned above, a precise determination of the total mass of
distant luminous clusters like MS1054-03 is crucial for the
determination of $\Omega_0$ based on cluster abundances at high
redshifts.  Unfortunately there are still important uncertainties on the
total mass estimate: statistical errors due to the errors on the
temperature measurement but also possible systematic errors if the
temperature is a biased estimator of the cluster mass.  In particular
the gas temperature might differ from the true virial temperature
if the cluster were not in hydrostatic equilibrium.

By considering the additional information on the gas mass, we can
however perform a consistency check on the total mass estimate.  From
a study of the intrinsic dispersion in the $L_{\rm X}$--$T$ relation,
Arnaud \& Evrard (1999) showed that the fractional variations of
$\fgas$ at fixed cluster mass is very small.  Since it is unlikely
that the gas mass fraction evolves with redshift, the derived gas mass
fraction of MS1054-03 should be consistent with the typical value of hot
nearby clusters.

As a reference we first consider the best fit ASCA temperature
(k$T=12.3~\kev$) and a $\Omega_0=1$ universe.  The corresponding gas mass
fraction of MS1054-03 is significantly smaller (at the 95$\%$
confidence level) than the value found by Arnaud \& Evrard (1999) for
hot nearby clusters, $\fgas = 20.1 \pm 1.6\ \%$.  Obviously MS1054-03
might be a true outlier.  Such outliers are rare but some, such as
A1060, a cluster of exceptional low gas content in the local Universe
(Arnaud \& Evrard 1999), do exist.  If this is not the case for MS1054, the 
low derived
$\fgas$ value suggests that the actual virial temperature is lower
than $12.3~\kev$ and/or that the adopted cosmology is wrong.  We
examine each possibility in turn.

The derived gas mass fraction depends on the assumed values of the
cosmological parameters $\Omega_0$ and $\Lambda$.  In the BM approach,
$\fgas \propto d_{\rm A}^{3/2}$, where $d_{\rm A}$ is the angular
distance (Pen 1997).  In the VT approach adopted here the dependence of
$\fgas$ is slightly different.  $\fgas$ is both dependent on the
geometry factor (the variation of $d_{\rm A}$) and on the
normalization of the $M_{\rm V}({\rm or} R_{\rm V})$--$T$ relation.
A detailed derivation of this dependence and necessary equations to
compute the variation of $\fgas$ with $\Omega_0$ are given in the
Appendix.  The variation of the gas mass fraction of MS1054-03 with
$\Omega_0$ is plotted on Fig.~5 for an open Universe
( $\Omega<1$, $\Lambda=0$) and a flat Universe ($\Omega_{0}+\Lambda=1$).  For 
the
best 1--D(2--D) \betamodel, the estimated gas mass fraction of
MS1054-03 would be a factor of 1.2(1.3) larger if $\Omega_0=0.3$,
$\Lambda=0.0$ and 1.4(1.5) times larger if
$\Omega_0=0.3$,$\Lambda=0.7$.  The gas mass fraction of MS1054-03
would in the later case be perfectly consistent with the local value,
which would itself remain essentially unchanged ($< 5\%$ increase).

The second possibility is that the virial temperature, and thus the
virial mass, are smaller than given by the best fit ASCA value.
The lower limit on the ASCA temperature (k$T=10.1~\kev$) yields a gas
mass fraction marginally consistent with the local value.  If the
virial temperature were actually even  somewhat lower, k$T=8~\kev$, the virial
mass would decrease to $0.84 \times 10^{15}\msun$, the virial radius
to 1.3 Mpc and the gas mass within that radius, $1.7 \times
10^{15}~\msun$, would reach $20\%$ of the virial mass and be  a perfect match
with the local value.  A virial temperature of $8 \kev$ is formally
excluded by ASCA measurements at the $90\%$ confidence level but would
be in better agreement with the measured velocity dispersion of
$\sigma = 1170 \pm 150$~kms/s (Tran \etal 1999).  In that case
MS1054-03 would fit perfectly in the $\sigma-T$ relation established
for $z=0.19-0.55$ clusters and the virial relation (Fig.~3 of Tran
\etal 1999).  The temperature would also be more consistent with the measured
bolometric luminosity.  A possible explanation for the measured
temperature being significantly higher than the virial temperature
(apart from contamination by a hard source like an
absorbed AGN) would be the presence of shock waves in the gas if the
cluster were indeed undergoing a recent merger.

\section{Conclusion}

Our wavelet analysis finds evidence for two components in the X--ray
image of MS1054-03: a main diffuse component, with emission peaking at
$20\arcsec$ from the brightest cluster member, and a compact 
substructure in the
west, which coincides with the X-ray maximum in the ROSAT image.  We
emphasize that this brightest peak is not the centroid of the cluster.
Indeed, when the western structure is excluded, the cluster emission
is fitted well by a classical \betamodel, centered within $20\arcsec$
of the central cD galaxy.  This indicates that we are then
identifying the main cluster component.

Unlike previous attempts (D98, Ebeling \etal 1999), we obtained a good
fit to the data with a \betamodel (low $\chi^2$ value) and derived
well constrained parameters.  This is a natural consequence of the
higher S/N ratio of our data (optimum channel selection and longer
exposure time) and our identification and removal of the western structure 
from our fits.

There are indications from the X--ray data that the cluster is not
fully relaxed.  The core radius ($\sim 400$~kpc) and ellipticity are
relatively high and the western substructure could be an in-falling
group.  This is not unusual, such mergers do exist in the nearby
Universe.  Actually MS1054-03 appears very similar to A521 at z=0.27 (Arnaud
\etal 2000), where the brightest peak is also associated with the
subcluster.

The gas mass fraction derived for the best fit ASCA temperature of
k$T=12.3~\kev$ is only consistent with the local value if $\Omega_{0}
< 1$, a flat $\Lambda$--dominated Universe  being favored.  The
local value can be matched as well for $\Omega_{0} = 1$, provided
that the actual virial temperature is close to the lower ASCA limit
($\sim 10~\kev$), with an even lower value of $8~\kev$ giving the best
match.  To decide between the two options requires a systematic
analysis of a large sample of distant clusters. MS1054-03 would appear
as an outlier in the second case (low virial temperature)
and not in the first case (flat $\Lambda$ dominated Universe).  If the
cluster's actual mass is indeed lower than previously estimated, this
might have consequences for the measurements of $\Omega_0$ based on the
abundance of massive clusters at high redshift.

Finally we want to stress that the current data allow us to constrain
the physics of MS1054 only with large error bars.  Therefore better
data, in particular spectroscopic data from {\it Chandra} and {\it
XMM}, are strongly needed.

\section*{Acknowledgments}
We want to thank Isabella Gioia and Megan Donahue for providing the
optical image (see Fig.~2).  We are grateful for discussions with
Megan Donahue, Marshall Joy, Sandeep Patel and Jack Hughes.

\section*{APPENDIX}
\appendix

\def \msol      {{\rm\ M}_\odot}

In this Appendix we consider the gas mass fraction derived from X-ray
data  in the VT approach (see Sect.\ref{sec:mass}) and examine how it
depends on the assumed cosmological parameters, $\Omega_{0}$ and
$\Lambda$. \\

The total mass  $M_{\rm V}$ is estimated from the measured temperature and the
mass-temperature relation derived from the virial theorem at given
density contrast:
\begin{equation}
M_{\rm V}   \propto   \left(\frac{\Delta_{\rm
c}(\Omega(z),\Lambda)\Omega_{0}}{\Omega(z)}\right)^{-1/2}\
(1+z)^{-3/2}\ \left({\rm k} T\right)^{3/2}
\label{equ:mt}
\end{equation}
It depends on $\Omega_{0}$ and $\Lambda$, via the density contrast,
$\Delta_{\rm c}(\Omega(z),\Lambda)$, and the  density
parameter of the universe at redshift z, $\Omega(z)$. The analytical 
expression of $\Delta_c$ and $\Lambda(z)$
can be
found in Bryan \& Norman (1998).  The corresponding virial radius
scales as:
\begin{equation}
R_{\rm V}  \propto  \left(\frac{\Delta_{\rm
c}(\Omega(z),\Lambda)\Omega_{0}}{\Omega(z)}\right)^{-1/2}\
(1+z)^{-3/2}\ \left({\rm k} T\right)^{1/2}
\label{equ:rt}
\end{equation} \\

We assume that the gas density $n_{\rm g}(r)$, follows a
$\beta$--model: $n_{\rm g}(r) = n_{\rm g,0}\left[1 + (r/r_{\rm
c})^{2}\right]^{-3\beta/2}$.  X-ray imaging data provides the central
surface brightness $S_{0}$, the slope parameter $\beta$ and the
angular core radius $\theta_{c}=r_{\rm c}/d_{\rm A}$, where $d_{A}$ is
the angular distance.

The emission measure
along the line of sight through the cluster center can be derived from $S_{0}$,
independently of any cosmological parameters :
\begin{equation} EM_{0} =
\frac{4~\pi~(1+z)^{4}~S_{0}}{\Lambda(T,z)}
\label{equ:em0}\end{equation}  where
$\Lambda(T,z)$ is the emissivity in the detector band, taking into account
the interstellar absorption and the instrument spectral response.

Assuming that the X-ray atmosphere extends up to the virial
radius, the central emission measure along the line of sight is linked
to the gas density by $EM_{0} \propto \int_{0}^{R_{\rm V}} n_{\rm
g}^{2}(r)~dr $, whereas the gas mass within $R_{V}$ is $M_{\rm gas}
\propto \int_{0}^{R_{\rm V}} n_{\rm g}(r)~r^{2}~dr$.
By combining the  two expressions, we can derive an expression for the  gas
mass that varies as:
\begin{equation}
M_{\rm gas}    \alpha    R_{\rm V}^{5/2} \sqrt{EM_{0}}~Q(\beta,x_{\rm c})
\label{equ:mgas}
\end{equation}
where we have introduced the form factor $Q(\beta,x_{\rm c}) =
\langle{n_{\rm g}}\rangle/\sqrt{\langle{n_{\rm
g}^{2}}\rangle_{los}}$. Here $\langle{n_{\rm g}}\rangle$ is the average
gas density within $R_{\rm V}$ and $\langle{n_{\rm g}^{2}}\rangle_{los}$
is the average along the line--of--sight passing through the cluster
center. For the $\beta$--model, this form factor depends on $\beta$ and
the scaled core
radius $x_{c}=r_{c}/R_{V}$:
\begin{eqnarray}
Q(\beta,x_{\rm c})&= & \frac{3 \int_{0}^{1}
\left[1 +
(x/x_{c})^{2}\right]^{-3\beta/2}~x^{2}~dx}{\sqrt{\int_{0}^{1}
\left[1 +
(x/x_{c})^{2}\right]^{-3\beta}~dx}} =  
\frac{\mbox{BETACF}(\frac{3}{2},\frac{3\beta}{2},\frac{1}{1+x_c^2})}
{\sqrt{\mbox{BETACF}(\frac{1}{2},3\beta-\frac{1}{2},\frac{1}{1+x_c^2})}}
\end{eqnarray}\\

BETACF is the continued fraction entering the expression of the incomplete Beta
 function $B_x(a,b)$ (see Press et al. 1986) with:

\begin{equation}
B_x(a,b) = \frac{x^a(1-x)^b}{a} \mbox{BETACF}(x,a,b)
\end{equation}

From Eq.\ref{equ:mt}, Eq.\ref{equ:rt}, Eq.\ref{equ:em0} and
Eq.\ref{equ:mgas}, the gas mass fraction, estimated from given X--ray
data, depends on the assumed cosmological parameters $\Omega_{0}$ and
$\Lambda$ as:
\begin{eqnarray}
f_{\rm gas}&\alpha & R_{\rm V}^{3/2} Q(\beta, \theta_{\rm c}d_{\rm
A}/R_{\rm V})
\end{eqnarray}
where $R_{\rm V} \propto \left(\Delta_{\rm
c}(\Omega(z),\Lambda)\Omega_{0}/\Omega(z)\right)^{-1/2}$ and $d_{\rm
A}$ is the angular distance.  Note that the variation with the
cosmological parameters is not the same for all clusters.  It depends
on the cluster temperature and shape (via the $\theta_{\rm c}d_{\rm
A}/R_{\rm V}$ factor).

The variation of $f_{\rm gas}$ with $\Omega_{0}$, for an open Universe
($\Lambda=0$) and a flat Universe ($\Omega_{0}+\Lambda=1$) is plotted
on Fig.~5 for the X--ray best fit parameters of MS1054-03
(${\rm k}T=12.3~{\rm keV}, \beta=0.96, \theta_{\rm c}=0.89\arcmin$).


\newpage

 \begin{deluxetable}{cccccccc}
 \tablecaption{Isothermal \betamodel fit results}
\startdata
 	& $S_0$	& $r_{c1}$	& $r_{c2}$	& $\beta$ &	$x_0$ &
	$y_0$  & $B$ \\
 	& (ct/s/arcmin$^2$) &	(arcmin) & (arcmin)	& &
	(J2000) & (J2000)
 & (ct/s/arcmin$^2$)\\
\tableline
  1-d & $(5.4\pm0.6)~10^{-3}$	& $0.89_{-0.21}^{+0.37}$ &	&
$0.96_{-0.22}^{+0.48}$ &		 & &   $(3.07\pm0.02)~10^{-3}$\\
  2-d & $(6.0\pm 0.4)~10^{-3}$ & $0.85 \pm 0.22$ &  $ 0.60 \pm 0.13$	&
$0.73\pm0.18$ & $10^h56^m58.6^s$	 &
  $-03^d37^m36^s$ & $(3.00 \pm 0.03)~10^{-3}$ \\
\enddata
 \label{tab:beta2}
 \end{deluxetable}

\begin{deluxetable}{llll}
\tablecaption{Mass estimates}
\startdata
  Method & Radius &  $M_{\rm gas}$ & $M_{\rm tot}$ \\
		 & (Mpc) & $(10^{14}{\rm h}_{50}^{-5/2}$~\msun) &
		 $(10^{15}$~\msun) \\
\tableline
  1--D BM & 1. & $1.11\pm 0.05$ & $1.1_{-0.2}^{+0.3}~{\rm h}_{50}^{-5/2}$ \\
   2--D BM & 1. & $1.27$ & $0.89$ \\
\tableline
  1--D BM & 1.65 & $1.9\pm 0.3$ & $2.0_{-0.4}^{+0.8}~{\rm h}_{50}^{-5/2}$ \\
   2--D BM & 1.65 & $2.5$ & $1.6$ \\
\tableline
 VT  & 1.65 &  & $1.6~~~~~~~{\rm h}_{50}^{-1}$\\
\enddata
 \label{tab:mass}
\end{deluxetable}


\begin{figure}
 \epsfxsize=1.\textwidth
 \epsfbox{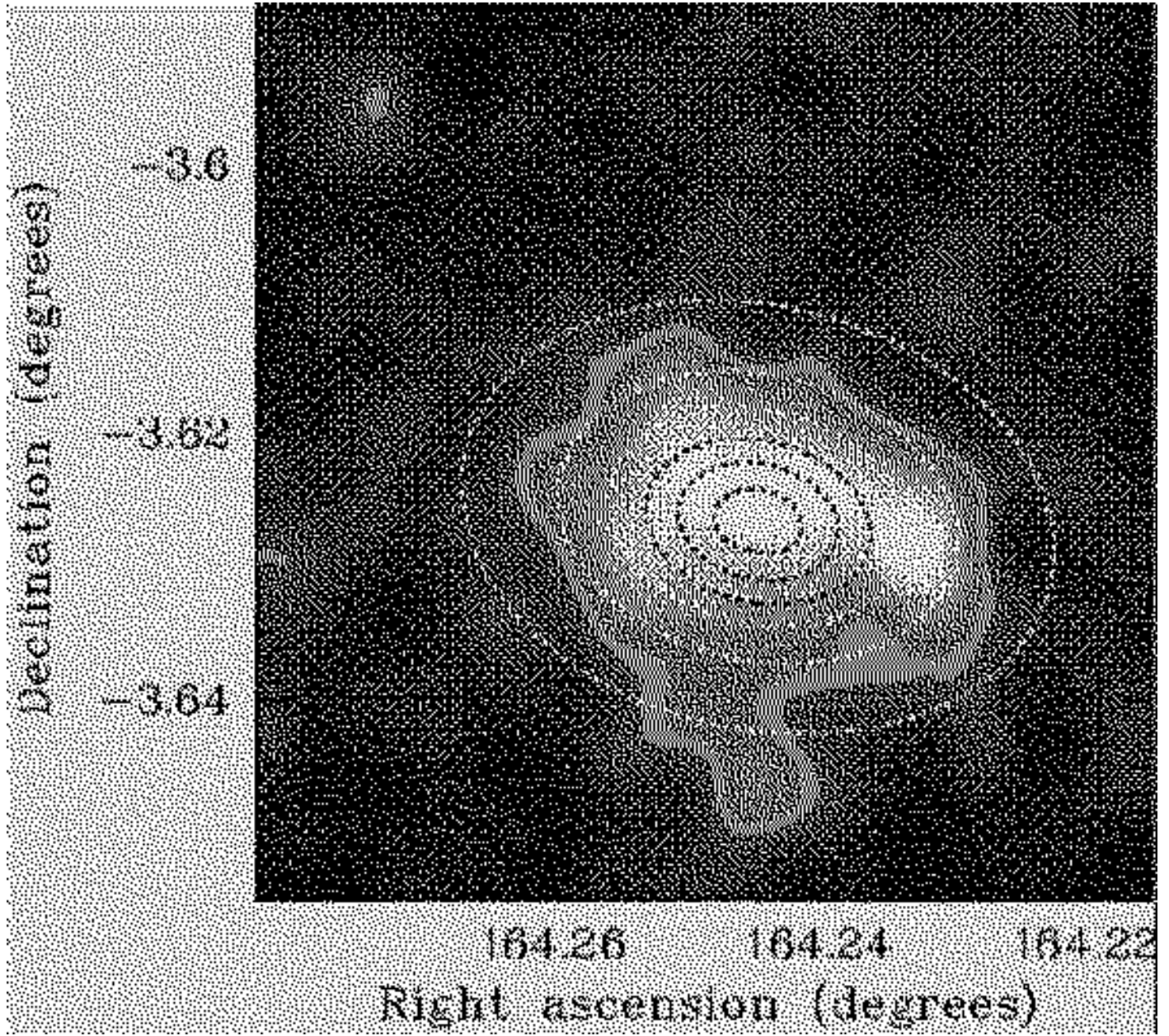}
 \caption[]{The grey scale image shows the ROSAT/HRI data after
 applying a Gauss filter of $\sigma=8\arcsec$.  The contours show the
 best fit elliptical cluster \betamodel ($\beta=0.73$, see also
 Tab.\ref{tab:beta2}).  The spacing of the contours is linear.  The
lowest contour is at $9.47~10^{-4}{\rm /s/arcmin^{2}}$ with similar
spacing between two contours.
 }
 \label{fig:ima}
\end{figure}


\begin{figure}
 \epsfxsize=1.\textwidth
\epsfbox{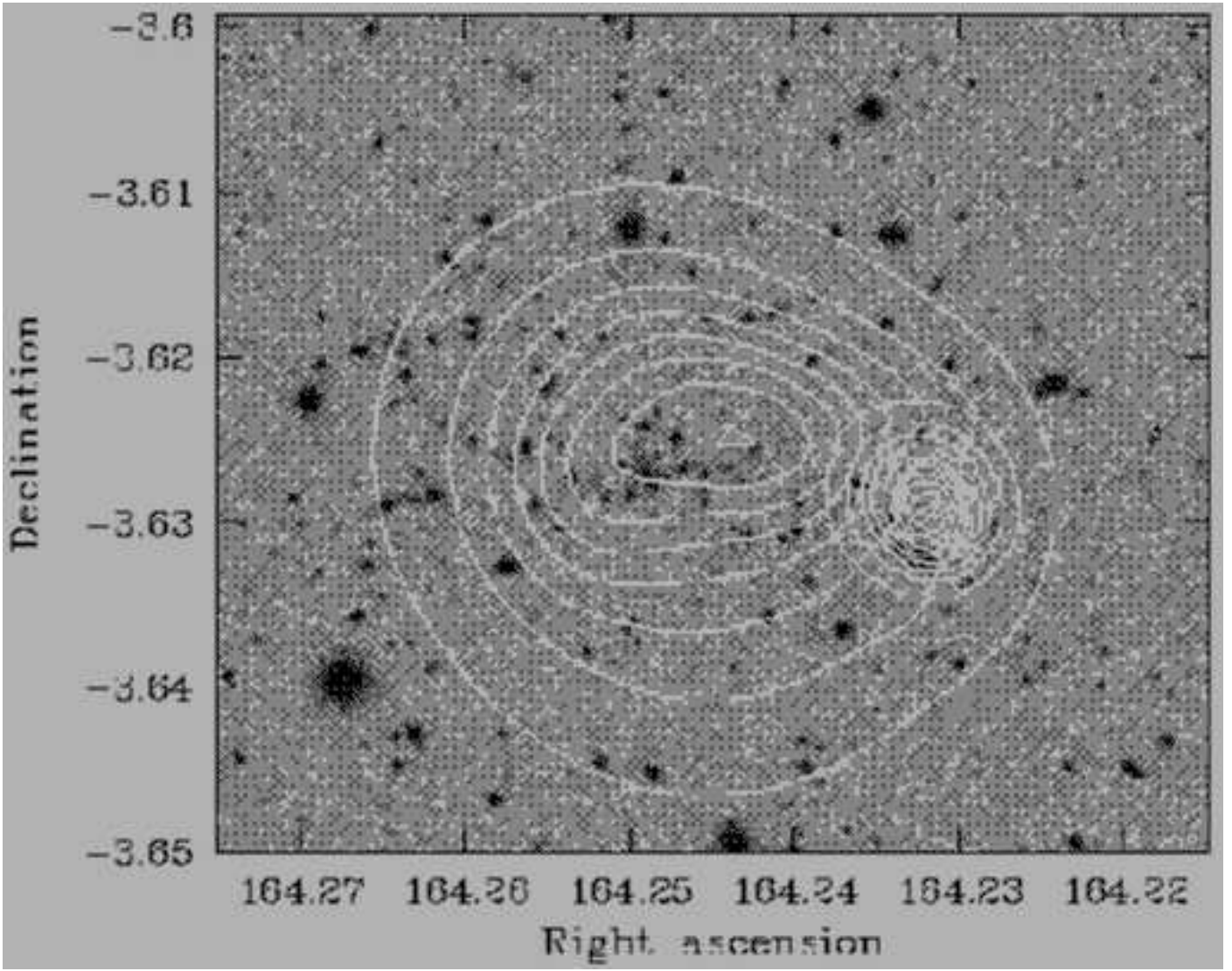}
\caption[]{ Restored
 X-ray image from the wavelet analysis of the HRI raw image.  The
 iso-intensity contours of the reconstructed main cluster and of the
 compact sub-substructure (in the west) are overlayed on the optical
 image (D98).  The isocontours are linearly spaced by $6.1~10^{-4}~{\rm
 ct/s/arcmin^{2}}$, the value of the
 lowest contour.}
 \label{fig:imawv}
 \end{figure}

\vfill
\newpage

\begin{figure}
\begin{center}
\epsfxsize=1\textwidth
 \epsfbox{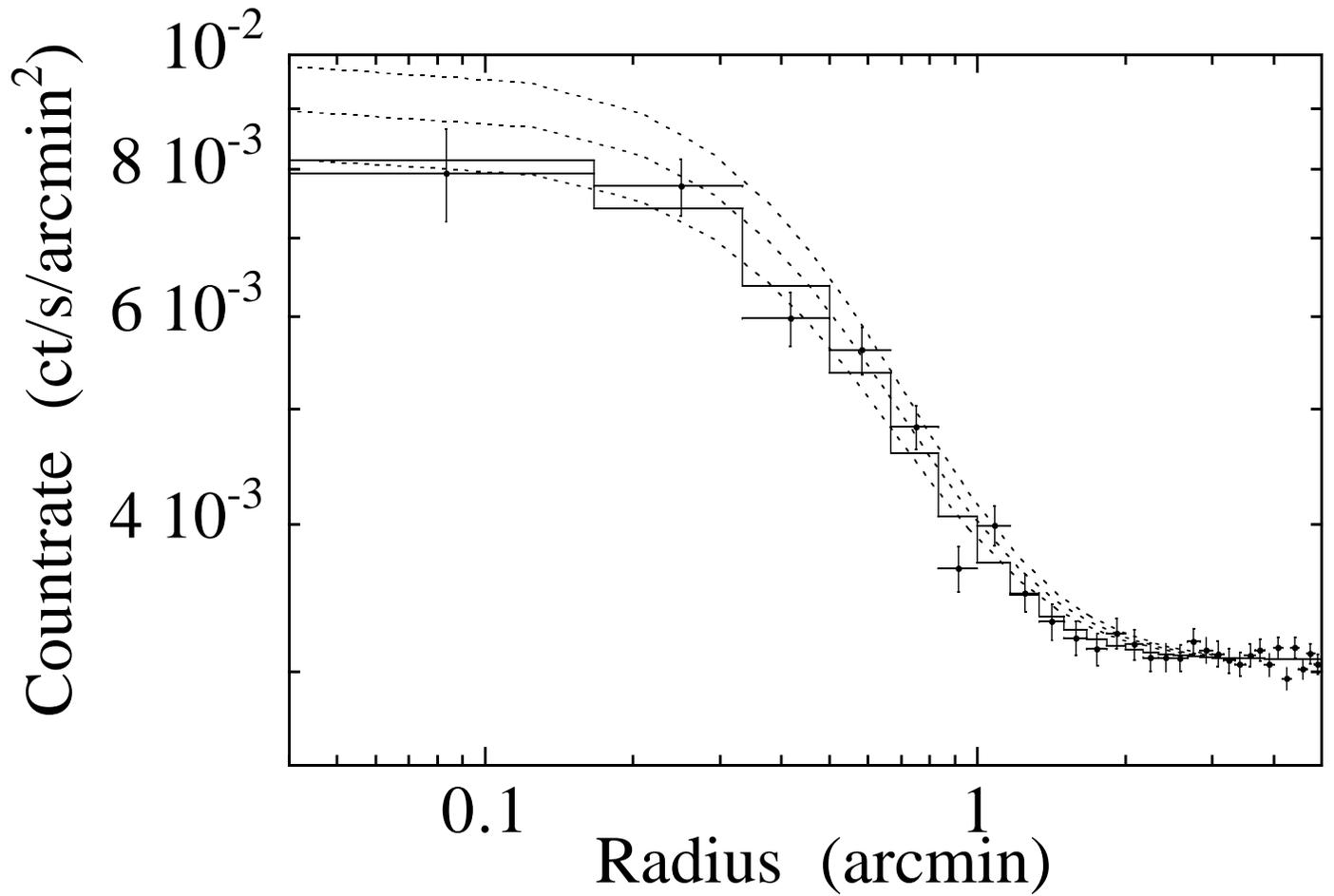}
\figcaption{Surface brightness profile of MS1054.  The crosses show the
data with 1$\sigma$ error bars.  The full line is the best fit 1--D
\betamodel.  The dotted curves are the profiles of the best fit 2--D
elliptical model $\pm 2\sigma$ errors on the central surface
brightness ($\beta=0.73$).}
\label{fig:surf}
\end{center}
\end{figure}

\vfill
\newpage

\begin{figure}
\begin{center}
 \epsfxsize=0.98\textwidth 
\epsfbox{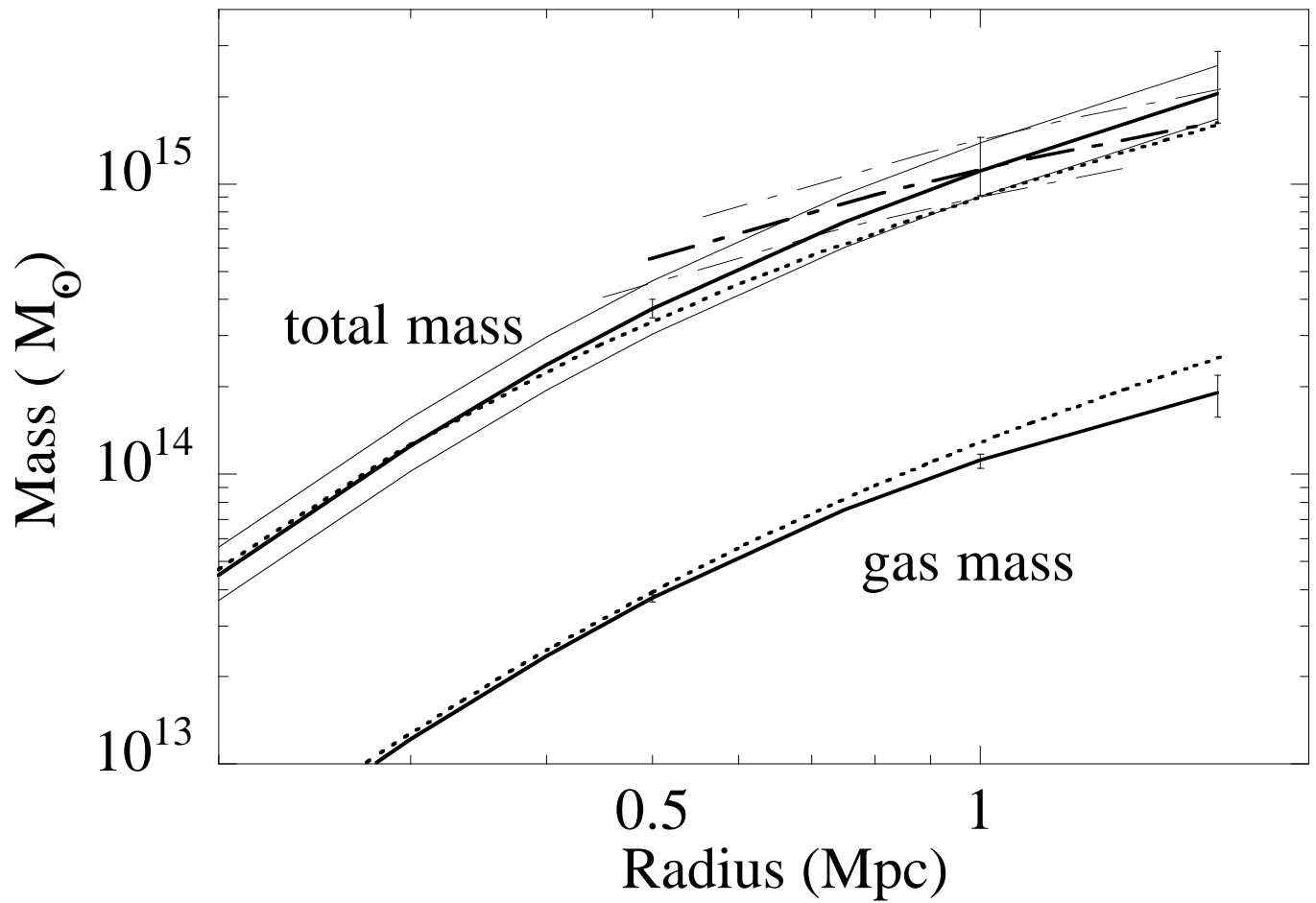} 
\figcaption{Total and
 gas mass profiles of MS1054.  The error bars correspond to the error
 on the imaging data.  Full lines and dashed lines are the results of
 the 1--D and 2--D isothermal \betamodel fit respectively.  Dashed-dotted 
lines: virial mass from numerical simulations at this redshift
 (Evrard \etal 1996). The heavy lines are for k$T=12.3 \kev$.  The
 thin lines correspond to the lower (k$T=10.1 \kev$) and upper limit
 (k$T=15.4 \kev$) of the ASCA temperature measurement.}
\label{fig:mass}
\end{center}
\end{figure}

\vfill
\newpage

\begin{figure}
\begin{center}
 \epsfxsize=0.98\textwidth 
\epsfbox{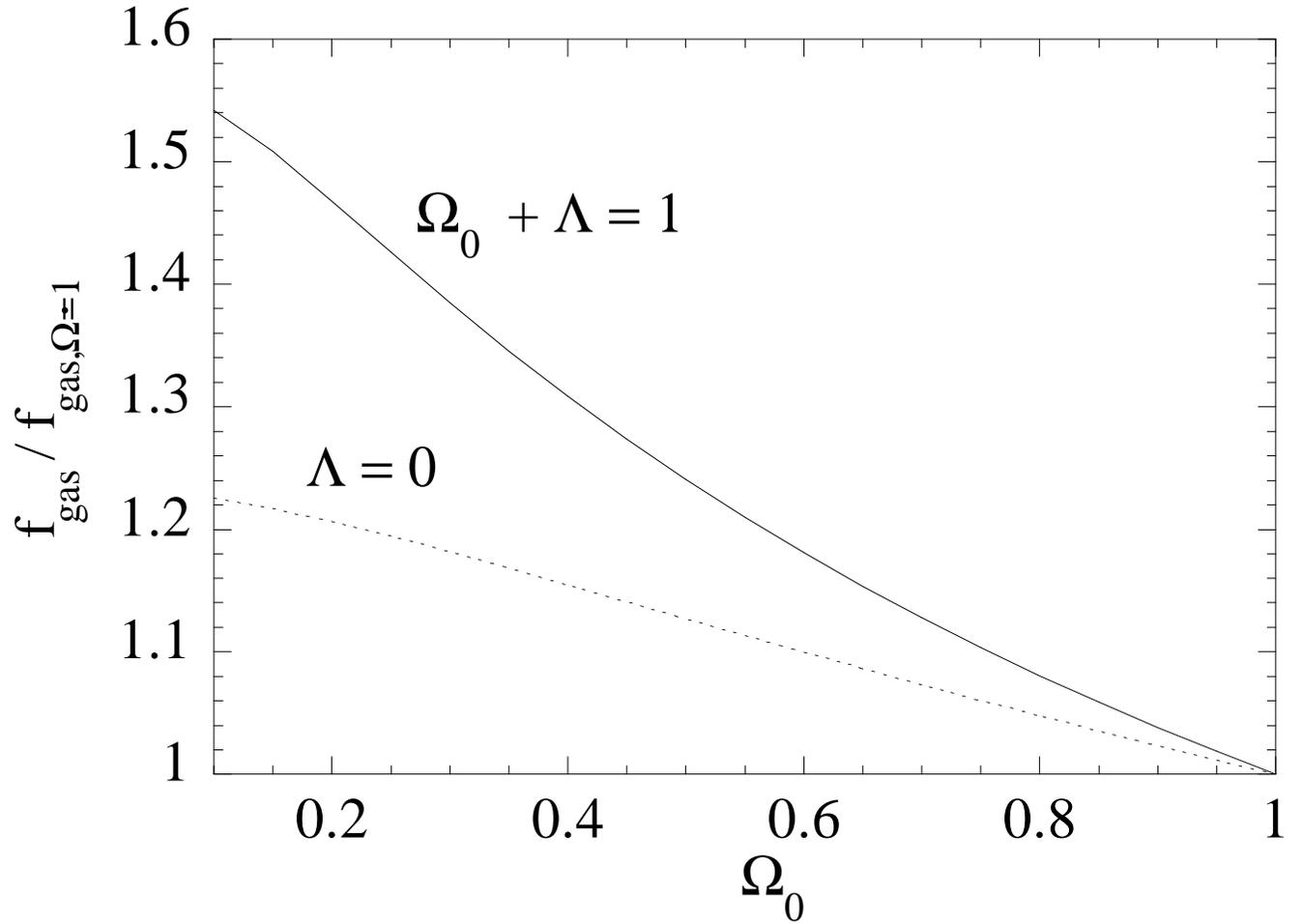}
 \figcaption{Variation of the estimated gas mass fraction $f_{\rm gas}$
 of
 MS1054-03 with the assumed cosmology.  $f_{\rm gas}$ is normalized to the
value
 obtained for $\Omega_{0}=1$. Full line: variation with $\Omega_{0}$
 for an open Universe ($\Lambda=0$). Dotted line: flat $\Lambda$
 dominated Universe.  The X--ray parameters are fixed to their best
 fit values (${\rm k}T=12.3~{\kev}, \beta=0.96, \theta_{\rm
 c}=0.89\arcmin$).}
\label{fig:cosmo}
\end{center}
\end{figure}

\end{document}